# PARALLEL SIMULATION OF THE MAGNETIC MOMENT REVERSAL WITHIN THE $\varphi_0$-JOSEPHSON JUNCTION MODEL

M. Bashashin[a,b,*], E. Zemlyanaya[a,b], I. Rahmonov[a,b,c]

[a]*Joint Institute for Nuclear Research, Joliot-Curie str. 6, 141980 Dubna, Russia*

[b]*Dubna State University, University str. 19, 141980 Dubna, Russia*

*e-mail:* bashashinmv@jinr.ru

[c]*Moscow Institute of Physics and Technology, 141700, Dolgoprudny, Russia*

**Abstract** — Periodic structure of the magnetization reversal domains is studied within the superconductor–ferromagnetic–superconductor $\varphi_0$-junction model. The model is described by the Cauchy problem for the system of nonlinear ordinary equations which is numerically solved by means of the 2-step Gauss–Legendre method. Two versions of parallel implementation on the basis of MPI and OpenMP techniques have been developed. Efficiency of both versions is confirmed by test calculations. An effect of frequency of ferromagnetic resonance on the configuration MR domains has been investigated. The calculations have been performed at the HybriLIT Platform of the JINR Multifunctional information and computing complex.

## INTRODUCTION

In the superconductor–ferromagnetic–superconductor structures, the spin-orbit coupling in ferromagnetic layer without inversion symmetry provides a mechanism for a direct (linear) coupling between the magnetic moment and the superconducting current [1]. Such Josephson junctions are called $\varphi_0$-junction. A perspective to control magnetic properties of $\varphi_0$-junctions by means of the superconducting current as well as effect of magnetic dynamics on the superconducting current is of great interest from the viewpoint of practical applications in nanoelectronic devises [1-3]. One of the important directions in this field is a study of phenomenon of magnetization reversal (MR) in $\varphi_0$-junction [4]. Different scenarios for the MR realization and the periodic structure of MR intervals are demonstrated in [5]. This contribution aims to study an influence of frequency of ferromagnetic resonance on the configuration MR domains. The basic formulae of the mathematical model are given. Numerical approach including the parallel implementation is briefly described. Results of numerical simulations of the MR phenomenon depending on the parameters of model are presented. Also, the results of test calculations demonstrating effect of parallel implementation on the basis of MPI and OpenMP techniques are given.

## THEORETICAL MODEL

We use the theoretical framework as in [4]. The model of the superconductor–ferromagnetic–superconductor $\varphi_0$-Josephson junction is based on the Landau–Lifshitz–Gilbert equation which describes the dynamics of the magnetization vector $\vec{m}$ in ferromagnetic layer in the $\varphi_0$-junction:

$$\frac{d\vec{m}}{dt} = -\frac{\omega_F}{1+\vec{m}^2\alpha^2}\{[\vec{m}\times\vec{H}] + \alpha[\vec{m}(\vec{m}\vec{H}) - \vec{H}\vec{m}^2]\}, \quad (1)$$

where $\alpha$ is damping parameter, $\omega_F$ is normalized frequency of ferromagnetic resonance, $\vec{H}$ is an effective magnetic field vector with the components

$$\begin{cases} H_x = 0 \\ H_y = Gr\sin(\varphi(t) - rm_y(t)), \\ H_z = m_z(t) \end{cases} \quad (2)$$

$G$ – relation of Josephson energy to energy of magnetic anisotropy, $r$ – the spin-orbit coupling parameter, $m_{x,y,z}$ corresponds the $x,y,z$-component of magnetization vector $\vec{m}$. The Josephson phase difference $\varphi$ can be found as follows:

$$\frac{d\varphi}{dt} = I_{pulse}(t) - \sin(\varphi - rm_y), \quad (3)$$

where the pulse current $I_{pulse}$ is given by

$$I_{pulse} = \begin{cases} A_s, & t \in [t_0 - 1/2\Delta t, t_0 + 1/2\Delta t] \\ 0, & \text{otherwise} \end{cases}. \quad (4)$$

Here $A_s$ is the amplitude of the pulse current, and $\Delta t$ is the time interval, in which the pulse current is applied. The system of equations (1) with effective field (2), (3) and with the pulse current (4) describes the dynamics of the $\varphi_0$-junction. The initial conditions are following:

$$m_x(0) = 0, m_y(0) = 0, m_z(0) = 1, \varphi_0 = 0. \quad (5)$$

Problem (1-5) is a Cauchy problem for a system of ordinary differential equations. Numerical solution is based on the implicit two-step Gauss–Legendre method which was shown in [6] to provide more accurate calculations in comparison with the explicit Runge–Kutta scheme.

## RESULTS OF NUMERICAL SIMULATIONS

Magnetic reversal is an effect when the component $m_z(t)$ changes its initial value $m_z(0)=+1$ and stabilizes on the value $m_z(T_{max})=-1$ for sufficiently large $T_{max}$. Numerical study is reduced to massive numerical simulations the Cauchy problem (1-5) with varied parameters of model. In [5], the results of such simulations at the planes of parameters $(G, \alpha)$ and $(G, r)$ are presented while all other parameters have the following values: at $A_s = 1.5$; $r = 0.1$; $t_0 = 25$; $\Delta t = 6$. Simulations were carried out with the time stepsize $h_t=0.01$ while $t \leq T_{max}=1000$. MR was indicated using the condition $|m_z(T_{max})+1|<\varepsilon$, where $\varepsilon=0.0001$. It was shown in [5] that the MR realization is periodic when the value of $G$ changes, i.e. there are intervals on $G$ where MR occurs and intervals where MR is absent. The purpose of this work is to demonstrate a dependence of configuration of the MR realization intervals on the parameter $\omega_F$ (normalized frequency of ferromagnetic resonance). To this end, massive numerical simulations were carried out on the $(G, \alpha)$ and $(G, r)$ planes in order to reveal the domains where the MR takes place. Three values of $\omega_F$ were used in calculations: $\omega_F=0.1$, 1, and 10. The results are presented in Figures 1 and 2.

Figure 1 demonstrates the MR domains at the $(G, \alpha)$-plane. Colored areas correspond the pairs of parameters $G$ and $\alpha$ where the MR occurs. One sees that at small $\omega_F=0.1$ we obtained a solid domain of G between 0 and 60 where the MR takes place. In case $\omega_F=1$, the MR domains are in the form of wide bands alternating with areas where the MR does not occur. As for the case of large $\omega_F=10$ – the MR domains structure looks rather chaotic due to the very narrow alternating intervals of presence and absence of the magnetization reversal.

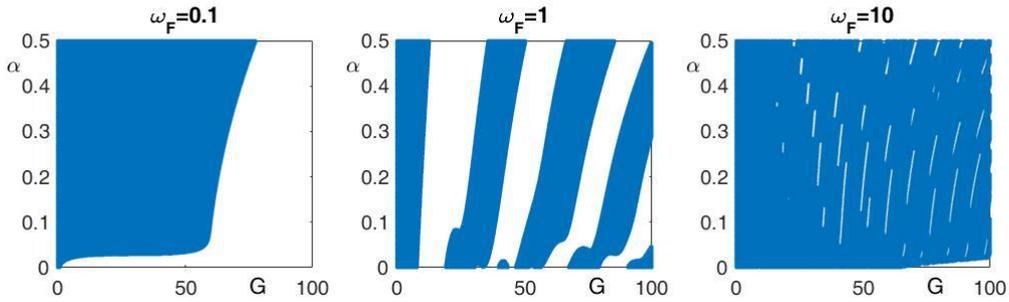

**Fig. 1.** Magnetization reversal domains at the $(G, \alpha)$-plane depending on value of $\omega_F$. The calculations were carried out with the $G$-stepsize $\Delta G=0.1$, $\alpha$-stepsize $\Delta\alpha=0.001$.

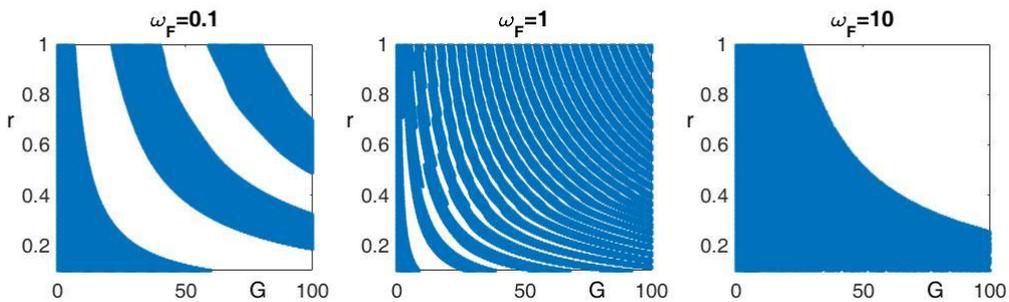

**Fig. 2.** Magnetization reversal domains at $(G,r)$-plane depending on value of $\omega_F$. The calculations were carried out with the $G$-stepsize $\Delta G = 0.1$, $r$-stepsize $\Delta r = 0.005$.

Figure 2 shows the MR domains at the $(G, r)$-plane. As in Figure 1, one observes that the magnetization reversal intervals (colored) alternate with the areas where the MR does not occur. Intervals of presence and absence of MR become narrow as $\omega_F$ grows from 0.1 to 10.

# EFFECT OF PARALLEL IMPLEMENTATION

Since numerical investigation of the MR effect requires massive calculations for numerical solution of the Cauchy problem (1-5) at a large number of pairs of parameters at ($G,\alpha$) and ($G,r$) planes, the parallel implementation have been developed to reduce the execution time and to accelerate the numerical study. In addition to the MPI parallel implementation presented in [7], we developed the OpenMP version to expand the possibilities of numerical research on different computing systems. In both MPI and OpenMP versions, the parallelism is based on a distribution of the points of ($G,\alpha$) or ($G,r$) plane between parallel processes. The calculations were made on two platforms of the Multifunctional Information Computing Centre [8] – the HybriLIT cluster and the Govorun supercomputer. The number P of parallel MPI processes and OpenMP threads was varied from 1 to 32. Both MPI and OpenMP implementations are confirmed to be quite effective and provide almost the same execution time in calculations at both HybriLIT cluster and Govorun supercomputer.

Figure 3 demonstrates that the execution time in test calculations at the ($G,r$) plane decreases hyperbolically with increasing a number P of parallel MPI processes and OpenMP threads. Weak oscillations of MPI-curves are explained by the strengthening of the interaction of parallel processes with increasing P (for details, see the recent work [9]).

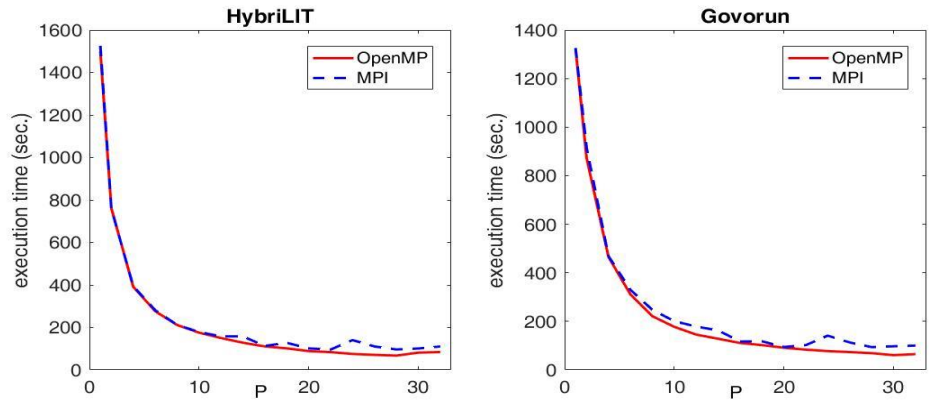

**Fig. 3.** The execution time of MPI and OpenMP calculations in dependence on P.

# SUMMARY

Massive numerical simulations in the wide range of parameters within the φ0-junction model allowed to obtained domains where the magnetic moment is reversed. An influence of the normalized frequency of the magnetic resonance $\omega_F$ on the periodic structure of MR domains has been studied. Test calculations confirm that both MPI and OpenMP parallel versions of the C++ computer code can provide the high-performance investigation of the MR phenomenon within the φ0-junction model.

The work was supported by the Russian Science Foundation grant № 18-71-10095 (https://rscf.ru/project/21-71-03009/).